\documentclass[twocolumn,aps,prl,showpacs,superscriptaddress,amsfonts,amssymb,amsmath]{revtex4}
\usepackage{graphicx}

\begin{document}

\title{Paired state in an integrable spin-1 boson model}

\author{Junpeng Cao}
\affiliation{Beijing National Laboratory for Condensed Matter
Physics, Institute of Physics, Chinese Academy of Sciences,
Beijing 100080, People's Republic of China}
\author{Yuzhu Jiang}
\affiliation{Beijing National Laboratory for Condensed Matter
Physics, Institute of Physics, Chinese Academy of Sciences,
Beijing 100080, People's Republic of China}
\author{Yupeng Wang*}
\affiliation{Beijing National Laboratory for Condensed Matter
Physics, Institute of Physics, Chinese Academy of Sciences,
Beijing 100080, People's Republic of China}
\affiliation{International Center for Quantum Structures, Chinese
Academy of Sciences, Beijing 100080, People's Republic of China}

\begin{abstract}
An exactly solvable model describing the low density limit of the
spin-1 bosons in a one-dimensional optical lattice is proposed.
The exact Bethe ansatz solution shows that the low energy physics
of this system is described by a quantum liquid of spin singlet
bound pairs. Motivated by the exact results, a mean-field approach
to the corresponding three-dimensional system is carried out.
Condensation of singlet pairs and coexistence with ordinary
Bose-Einstein condensation are predicted.
\end{abstract}


\maketitle

\section{Introduction}

Study on the trapped cold atoms opens the door for finding new
matter states which are usually unknown or even do not exist in
nature. Experimentally, the cold atom gas can be realized by means
of either magnetic or optical traps. With Feshbach resonance, the
scattering length and thus the couplings among atoms can be
manipulated experimentally. In addition, with laser beams, one can
confine particles in valleys of periodic potential of the optical
lattice. These experimental tools provide a platform to study
quite clean and controllable ``artificial condensed matter
systems". Moreover, particles with higher inner degrees of freedom
(hyperfine spin), which usually do not exist in conventional
condensed matters, can be prepared by catching several hyperfine
sublevels of atoms. Compared to spinless Bose gases, the
low-energy physics of these systems such as the spin
dynamics\cite{diener1,diener11,diener2} is much richer and may
show fascinating macroscopic quantum phenomena. For example, the
multi-component Bose-Einstein condensation (BEC) is realized in
$^{87}Rb$ \cite{1} and $^{23}Na$ \cite{2,3,4} gases with optical
traps. Both $^{87}Rb$ and $^{23}Na$ atoms have a hyperfine spin
$F=1$. The interaction among $^{87}Rb$ atoms is ferromagnetic,
which leads to a spin-polarized (ferromagnetic) ground state,
while the spin exchange interaction among the $^{23}Na$ atoms is
antiferromagnetic, leaving the ground state a spin singlet
condensate. In an optical lattice, the Mott phase of $F=1$ cold
atoms may exhibit rich magnetic structures. Nematic singlet
\cite{8} or dimerized \cite{9} ground state has been proposed.
Nevertheless, study on spinor cold atoms is still young and a
quite interesting issue \cite{10,11,12,13} in modern many body
physics.

In this Letter, we propose an exactly solvable model for $F=1$
bosonic cold atoms. The Bethe ansatz solution exactly shows that
atoms may form spin singlet pairs with a finite energy gap and the
low-energy physics is described by a quantum liquid of spin singlet
atom pairs. Based on the exact solution for the 1D model, an
appropriate mean-field theory is proposed to study the corresponding
3D systems. BCS-like pair condensation and coexistence with ordinary
BEC are found in the 3D model.

\section{The model}

\par
In an optical lattice, it has been proposed that the following boson
Hubbard model \cite{boson1,boson2} well describes the low-energy
physics of the spinor bosons:
\begin{eqnarray}
H=&-&t\sum_{<i,j>,s}(a_{i,s}^\dagger
a_{j,s}+h.c.)+\frac{U_0}2\sum_in_i(n_i-1)\nonumber\\
&+&\frac{U_2}2\sum_i({\bf S}_i^2-2n_i)-\mu\sum_in_i,
\end{eqnarray}
where $a_{i,s}^\dagger$ ($a_{i,s}$) is the creation (annihilation)
operator of atoms on site $i$ with spin index $s$, $n_i$ and ${\bf
S}_i$ are the particle number and spin operators, respectively;
$\mu$ is the chemical potential.
\par
Recently, tremendous experimental and theoretical progress has been
achieved in realization of one-dimensional (1D) cold atom systems
\cite{a,h,h1,b,t,bl} and 2D systems\cite{re}. The Mott phase diagram
of $F=1$ bosons in an optical lattice has been given in
ref.\cite{13}. In the metallic phase, it is known that the 1D
spinless bosonic atom gases are well described by Lieb-Liniger model
\cite{lieb,lieb1} and several physical properties based on
Lieb-Liniger's exact results have been derived\cite{fuchs,liebseir}.
However, results on 1D cold atoms with internal degrees of freedom
in the metallic phase are still rare. Generally speaking, a 1D
exactly solvable model not only gives the best understanding for the
corresponding universal class, but also provides some useful clues
for understanding three-dimensional (3D) systems.
\par
In this paper, instead of studying model (1), we consider the
following 1D Hamiltonian:
\begin{eqnarray}
H=-\sum_{i=1}^N \frac{\partial^2}{\partial x_i^2} +\sum_{i<j} [c_0 +
c_2{\bf S}_i \cdot {\bf S}_j]\delta(x_i-x_j), \label{h2}
\end{eqnarray}
where  ${\bf S}_i$ is the spin operator with $z$-components
$s=\uparrow, 0, \downarrow$; $c_0=(g_0+2g_2)/3$, $c_2=(g_2-g_0)/3$,
$g_S=4\pi \hbar^2 l_S/ M_b$, $M_b$ is the mass of boson and $l_S$ is
the $s$-wave scattering length in the total spin $S$ channel
\cite{diener1,diener11}.  In the second quantization form, we define
the particle creation (annihilation) operators as $a_s^\dagger(x)$
($a_s(x)$). Obviously, the model (2) is just the low density limit
of the boson Hubbard model (1). We note that two-particle scattering
processes keep the conservation of the total spin $S$ and therefore
the model possesses an $SU(2)$ invariance. Non-trivial scattering
occurs only in the $S=0$ and $S=2$ channels. In the $S=1$ channel
the wave-function is antisymmetric by exchanging  two particles and
the delta-function interaction is irrelevant. Especially in the
$S=0$ channel, a special scattering process
\begin{eqnarray}
a_{\uparrow}^\dagger(x)+a_{\downarrow}^\dagger(x)\to 2a_0^\dagger(x)
\end{eqnarray}
occurs, which makes the total particle number of an individual spin
component is no longer a good quantum number and breaks the $SU(3)$
invariance. It is easy to verify that the present model has the
following conserved quantities:
\begin{eqnarray}
N&=&\sum_s\int a_s^\dagger(x)a_s(x)dx,\nonumber\\
S^z&=&\int[a_{\uparrow}^\dagger(x)a_{\uparrow}(x)-a_{\downarrow}^\dagger(x)a_{\downarrow}(x)]dx,
\end{eqnarray}
where ${ N}$ and $S^z$ are the total particle number operator and
$z$-component of the total spin operator, respectively. Because of
the $SU(2)$ invariance of the Hamiltonian, there are also two other
good quantum numbers:
\begin{eqnarray}
S^+=\sqrt{2}\int[a_{\uparrow}^\dagger(x)a_0(x)+a_0^\dagger(x)a_{\downarrow}(x)]dx,\nonumber\\
S^-=\sqrt{2}\int[a_0^\dagger(x)a_{\uparrow}(x)+a_{\downarrow}^\dagger(x)a_0(x)]dx.
\end{eqnarray}
${S^z}$ and ${ S^\pm}$ form the generators of the $SU(2)$ algebra.
These three spin operators, combined with the five spin quadrupole
operators
\begin{eqnarray}
Q_0&=&\int[a^\dagger_\uparrow(x)a_{\uparrow}(x)
+a^\dagger_\downarrow(x)a_{\downarrow}(x)-2a_0^\dagger(x)a_0(x)]dx,\nonumber\\
Q_2&=&\int[a^\dagger_\uparrow(x)a_\downarrow(x)+a^\dagger_\downarrow(x)a_{\uparrow}(x)]dx,\nonumber\\
Q_{xy}&=&-i\int[a^\dagger_\uparrow(x)a_\downarrow(x)-a^\dagger_\downarrow(x)a_{\uparrow}(x)]dx,\\
Q_{xz}&=&\frac1{\sqrt{2}}\int[a_\uparrow^\dagger(x)a_0(x)-a_0^\dagger(x)a_\downarrow(x)+h.c.]dx,\nonumber\\
Q_{yz}&=&-\frac
i{\sqrt2}\int[a_\uparrow^\dagger(x)a_0(x)-a_0^\dagger(x)a_\downarrow(x)-h.c.]dx,\nonumber
\end{eqnarray}
form the basic representation of the $SU(3)$ algebra.

\section{Bethe Ansatz Solution}
\par
The pioneer work on the integrable models with internal degrees of
freedom was done by Yang\cite{yang,yang1} and followed by
Sutherland\cite{yang3}. There are two integrable lines for the model
(2). The first is the $c_2=0$ case, i.e., $SU(3)$-invariant case,
which has been solved by Sutherland\cite{yang3}. The second
integrable line is $c_0=c_2$, which has never been studied before
and is the main target of the present work. In the framework of
coordinate Bethe ansatz, the wave function of the system described
by a set of quasi-momenta $\{k_j\}$ can be written
as\cite{yang,yang1}
\begin{eqnarray}
&&\Psi(x_1s_1, \cdots,
x_Ns_N)=\sum_{Q,P}\theta(x_{Q_1}<\cdots<x_{Q_N})\nonumber \\
&& \quad\quad \times A_{s_1\cdots s_N} (Q,
P)e^{i\sum_{l=1}^{N}k_{P_l}x_{Q_l}},
\end{eqnarray}
where $Q=(Q_1, \cdots, Q_N)$ and $P=(P_1, \cdots, P_N)$ are the
permutations of the integers $1,\cdots, N$, $\theta(x_{Q_1}<\cdots
<x_{Q_N})=\theta( x_{Q_N}-x_{Q_{N-1}})\cdots \theta(
x_{Q_2}-x_{Q_1})$ and $\theta(x-y)$ is the step function. The wave
function is symmetric under permutating both the coordinates and the
spins of two atoms. The wave function is continuous but its
derivative jumps when two atoms touch. With the standard coordinate
Bethe ansatz procedure, we obtain the two-body scattering matrix for
$c_0=c_2=c$ as
\begin{eqnarray}
S_{ij}=\frac{k_i-k_j-ic}{k_i-k_j+ic}P_{ij}^0+P_{ij}^1+\frac{k_i-k_j+2ic}{k_i-k_j-2ic}P_{ij}^2,
\label{s1}
\end{eqnarray}
where $P_{ij}^S,{~~} S=0,1,2$ is the spin projection operator onto
the state of total spin $S$. The scattering matrix satisfies the
Yang-Baxter equation\cite{yang,yang1,yang3}
\begin{eqnarray}
&&S_{12}(k_1-k_2)S_{13}(k_1-k_3)S_{23}(k_2-k_3)\nonumber \\
&&\quad = S_{23}(k_2-k_3)S_{13}(k_1-k_3)S_{12}(k_1-k_2),
\end{eqnarray}
which ensures the integrability of the model (\ref{h2}) at
$c_0=c_2=c$. With the periodic boundary conditions of the wave
function, we obtain the following eigenvalue equations
\begin{eqnarray}
S_{jN}S_{jN-1}\cdots S_{jj+1}S_{jj-1}\cdots
S_{j1}e^{ik_jL}\xi_0=\xi_0, \label{ei}
\end{eqnarray}
where $\xi_0$ is the amplitude of initial state wave function. We
follow the algebraic Bethe ansatz method developed in
\cite{babujian,babujian1,schlottmann} to solve the above eigenvalue
problem. In fact, the $S$-matrix of the present model has the same
structure to that of the $R$-operator of the Takhtajan-Babujian
model\cite{babujian,babujian1}. In such a sense, the spin dynamics
of our model keeps some similarity to that of the Takhtajan-Babujian
spin chain. Firstly, we define the monodromy matrix as
\begin{eqnarray}
{\cal T_0}(\lambda)&=& S_{0j}S_{0N}S_{0N-1}\cdots
S_{0j+1}S_{0j-1}\cdots
S_{01}\nonumber \\
&=&\left(
\begin{array}{ccc}
A_1(\lambda) & B_1(\lambda) & B_2(\lambda) \\
C_1(\lambda) &A_2(\lambda) & B_3(\lambda) \\
C_2(\lambda) &C_3(\lambda) & A_3(\lambda)
\end{array}
\right),\label{m1}
\end{eqnarray}
where $S_{0l}\equiv S_{0l}(\lambda-k_l)$. The eigenvalue problem
(10) is therefore reduced to
\begin{eqnarray}
tr_0 {\cal T_0}(k_j)e^{ik_jL}\xi_0=\xi_0. \label{t1}
\end{eqnarray}
The monodromy matrix satisfies the Yang-Baxter relation
\begin{eqnarray}
S_{12}(\lambda-u) {\cal T}_1(\lambda) {\cal T}_2(u) ={\cal
T}_2(u){\cal T}_1(\lambda) S_{12}(\lambda-u). \label{ybe1}
\end{eqnarray}
Further, we define an auxiliary monodromy matrix as
\begin{eqnarray}
T(\lambda)&=&S_{0j}^{\sigma s}S_{0N}^{\sigma s}\cdots
S_{0j+1}^{\sigma s}S_{0j-1}^{\sigma s}S_{01}^{\sigma s}\nonumber \\
&=&\left(
\begin{array}{cc}
A(\lambda) & B(\lambda) \\
C(\lambda) & D(\lambda)
\end{array}
\right), \label{m2}
\end{eqnarray}
with
\begin{eqnarray}
S_{0l}^{\sigma s}(\lambda)=\frac{ \lambda-k_l -i\frac{1}{2}c -
ic{\bf \sigma}_0 \cdot {\bf S}_l }{\lambda-k_l+i\frac{3}{2}c}.
\end{eqnarray}
The monodromy matrices (\ref{m1}) and (\ref{m2}) satisfy the
Yang-Baxter relations
\begin{eqnarray}
S_{12}^{\sigma s}(\lambda-u) T_1(\lambda) {\cal T}_2(u) ={\cal
T}_2(u)T_1(\lambda) S_{12}^{\sigma s}(\lambda-u),
\nonumber\\
S_{12}^{\sigma\sigma}(\lambda-\mu)T_1(\lambda)T_2(\mu)
=T_2(\mu)T_1(\lambda)S_{12}^{\sigma\sigma}(\lambda-\mu),\label{ybe2}
\end{eqnarray}
with $S_{12}^{\sigma\sigma}(\lambda)=(\lambda-
ic)^{-1}(\lambda-ic/2-ic{\bf \sigma}_1\cdot{\bf \sigma}_2/2)$.
From Eq. (\ref{ybe2}) we obtain the following commutation
relations
\begin{eqnarray}
A_1(\lambda)B(u)=\frac{\lambda-u+i\frac{3}{2}c}{\lambda-u-i\frac{1}{2}c}
B(u)A_1(\lambda) \nonumber \\
- \frac{i\sqrt{2}c}{\lambda-u-i\frac{1}{2}c} B_1(\lambda)A(u),
\hspace{1cm}
\end{eqnarray}
\begin{eqnarray}
A_2(\lambda)B(u)=\frac{(\lambda-u+i\frac{3}{2}c)(\lambda-u-i\frac{3}{2}c)}
{(\lambda-u+i\frac{1}{2}c)(\lambda-u-i\frac{1}{2}c)}
B(u)A_2(\lambda)
\nonumber \\
 + \frac{i\sqrt{2}c}{\lambda-u-i\frac{1}{2}c} B_1(\lambda)D(u) -
\frac{i\sqrt{2}c}{\lambda-u+i\frac{1}{2}c}
B_3(\lambda)A(u) \nonumber \\
+ \frac{2ic}{(\lambda-u+i\frac{1}{2}c)(\lambda-u-i\frac{1}{2}c)}
B_2(\lambda)C(u),\hspace{1.55cm}
\end{eqnarray}
\begin{eqnarray}
A_3(\lambda)B(u)=\frac{\lambda-u-i\frac{3}{2}c}{\lambda-u+i\frac{1}{2}c}
B(u)A_3(\lambda)\nonumber \\
+ \frac{i\sqrt{2}c}{\lambda-u+i\frac{1}{2}c}
B_3(\lambda)D(u).\hspace{1cm}
\end{eqnarray}
Meanwhile, the commutation relations of $A(\lambda)$, $D(\lambda)$
and $B(\lambda)$ read
\begin{eqnarray}
A(\lambda)B(u)=\frac{\lambda-u+ic}{\lambda-u} B(u)A(\lambda)
-\frac{ic}{\lambda-u} B(\lambda)A(u), \\
D(\lambda)B(u)=\frac{\lambda-u-ic}{\lambda-u} B(u)D(\lambda)
+\frac{ic}{\lambda-u} B(\lambda)D(u).
\end{eqnarray}

The vacuum state of the system is defined as
$|\Omega\rangle=|\uparrow\rangle_1\otimes\cdots\otimes|\uparrow\rangle_N$.
It is a common eigenstate of $A_1(\lambda)$, $A_2(\lambda)$,
$A_3(\lambda)$, $A(\lambda)$ and $D(\lambda)$. The element
$C(\lambda)$ acting on the vacuum state gives zero. The element
$B(\lambda)$ acting on the vacuum state gives nonzero values and
thus can be regarded as generating operator of eigenstates
\begin{equation}
|\Psi\rangle =B(u_1)\cdots B(u_M)|\Omega\rangle. \label{st}
\end{equation}
$tr_0 {\cal T}(k_j)\equiv \sum_{n=1}^3 A_n(k_j)$ acting on the
assumed Bethe states (\ref{st}) gives two kinds of terms, i.e.,
wanted and unwanted terms. Putting the unwanted terms to be zero we
readily obtain the following Bethe ansatz equations for the
rapidities $\{k_j\}$,
\begin{eqnarray}
e^{ik_jL}=\prod_{l=1,l\neq j}^{N} \frac{k_j-k_l+2ic} {k_j-k_l-2ic}
\prod_{\alpha=1}^{M}\frac{k_j-\Lambda_{\alpha}-ic}
{k_j-\Lambda_{\alpha}+ic},  \label{bae1}
\\
\prod_{l=1}^{N} \frac{\Lambda_{\alpha}-k_l-ic}
{\Lambda_{\alpha}-k_l+ic}=- \prod_{\beta=1}^{M}
\frac{\Lambda_{\alpha}-\Lambda_{\beta}-ic
}{\Lambda_{\alpha}-\Lambda_{\beta}+ic }, \hspace{1cm}\label{bae2}
\end{eqnarray}
where $j,l=1, \cdots, N$, $\alpha,\beta=1,\cdots, M$, $M$ is the
number of flipped spins and $\{\Lambda_{\alpha}\}$ is the set of the
spin rapidities. The corresponding eigenvalue of the Hamiltonian (2)
reads
\begin{eqnarray}
E=\sum_{j=1}^N k_j^2.
\end{eqnarray}

\section{Thermodynamic limit}

Above we have confined the particles in a finite 1D box with
length $L$. Based on the solutions of the Bethe ansatz equations,
we can study the ground state and low-temperature properties of
the system in the thermodynamic limit $L\to\infty$, $N/L\to n$.
The solutions of the Bethe ansatz equations are a little bit
complicated. Besides real solutions of
$\{k_j\},{~}\{\Lambda_\alpha\}$, Eqs. (\ref{bae1}-\ref{bae2}) have
also complex solutions for both $c>0$ and $c<0$, which are usually
called as string solutions. For $c>0$, attractive interaction only
occurs in the $S=0$ channel. That means particles may form spin
singlet bound pairs. Generally, the complex solutions are
determined by the poles or zeros of the Bethe ansatz equations in
the thermodynamic limit. For example, if some $k_j$ in the upper
complex plane, the left side of Eq. (\ref{bae1}) tends to zero
when $L\to\infty$. Correspondingly, there must exist a
$\Lambda_\alpha$ satisfying $k_j-\Lambda_\alpha-ic\to 0$.
Furthermore, from Eq. (\ref{bae2}) we learn that there is another
$\Lambda_\beta$ with $\Lambda_\alpha-\Lambda_\beta+ic\to 0$. For
the complex conjugate invariance of the equations, we obtain the
simple
 conjugate
$k_j$-pair solutions
\begin{eqnarray}
k_j&=&K_j+ic/2+ o(e^{-\delta L}),\nonumber\\
k_j^{*}&=&K_j- ic/2 + o(e^{-\delta L}), \label{string1}
\end{eqnarray}
associated with $\Lambda$ 2-strings
\begin{eqnarray}
\Lambda_{j}&=& K_{j} + ic/2 + o(e^{-\delta^{\prime} L}),\nonumber\\
\Lambda_{j}^{*}&=& K_{j} - ic/2 + o(e^{-\delta^{\prime} L}),
\label{string2}
\end{eqnarray}
where $K_j$ is a real parameter, $\delta$ and $\delta^\prime$ are
some positive constants. We studied the 3 and 4-particle cases and
verified that Eq. (\ref{string1}) describes the only possible bound
state in the charge sector. In fact, no more than two atoms can form
a bound state because of the symmetry constraint of the wave
functions. In the thermodynamic limit, each bound pair contributes
bound energy $\Delta=c^2/2$. Therefore, the low energy physics of
the present system must be described by a quantum liquid of these
bound pairs. In the ground state, all particles form such kind of
bound pairs (Even $N$ is supposed. For odd $N$, there is a single
unpaired particle and the ground state is 3-fold degenerate).
Substituting these 2-string ansatz into Eqs. (\ref{bae1}-\ref{bae2})
and taking logarithm, we arrive at one set of reduced Bethe ansatz
equations
\begin{eqnarray}
K_jL=\pi I_{j}-\sum_{l=1,l\neq j}^{N}
\left[\arctan\left(\frac{2(K_j-K_l)}{3c}\right)\right. \quad \quad \nonumber \\
\left. +\arctan\left(\frac{K_j-K_l}{c}\right)-
\arctan\left(\frac{2(K_j-K_l)}{c}\right)\right], \label{bae3}
\end{eqnarray}
where $I_{j}$ is integer (half integer) for $N/2$ odd (even). The
ground state corresponds to a sequence of consecutive $I_j$'s
around zero symmetrically. In the thermodynamic limit,define
$\rho_0(K_j)=L^{-1} (I_{j+1}-I_j)/(K_{j+1}-K_j))$ as the density
of flipped spins. Taking derivative of Eq. (\ref{bae3}), we obtain
that the density distribution $\rho_0(K)$ in the ground state
satisfies the following integral equation
\begin{eqnarray}
\rho_0(K)=\frac{1}{\pi}+ \frac{1}{\pi}\int_{-Q}^{Q}\left[
\frac{6|c|}{9c^2+4(K-K^{\prime})^2} \right. \hspace{1.5cm} \nonumber \\
\left. +\frac{|c|}{c^2+(K-K^{\prime})^2}-
\frac{2|c|}{4c^2+(K-K^{\prime})^2}\right]\rho_0(K^{\prime})dK^{\prime},
\end{eqnarray}
where the pseudo Fermi point $Q$ is determined by
\begin{eqnarray}
n=2\int^Q_{-Q}\rho_0(K)dK. \end{eqnarray} The density of the ground
state energy reads
\begin{eqnarray}
\frac{E_0}{L}=\int_{-Q}^{Q}\left(
2K^2-\frac{c^2}{2}\right)\rho_0(K)dK.
\end{eqnarray}
Obviously, the ground state is a global spin singlet with
$S^z=N-M=0$. However, it is not an $SU(3)$ singlet state. In the
insulator phase of the boson-Hubbard model, a spin nematic state or
a spin quadrupole polarized state \cite{zhang} has been obtained. In
our case, the spin part of the wave function of each bound pair
takes the form $(|\uparrow,\downarrow
\rangle+|\downarrow,\uparrow\rangle-|0,0\rangle)/\sqrt{3}$. It can
be easily deduced that the expectation values of the quadrupole
momenta per unit length are $\langle Q_0\rangle=-n/2$ and $\langle
Q_\alpha \rangle=0$ for $\alpha\neq 0$. There is a finite energy gap
$\Delta=c^2/2$ for the spin excitations. The only basic gapless
excitation is in the charge sector. This can be realized by either
digging a hole in the $K$ pseudo Fermi sea or putting a particle
above the pseudo Fermi point (Note we treat an atom pair as a single
particle here). The excitation energy $\epsilon(K)$ of a hole or a
particle with quasi momentum $P(K)=\pi I(K)/L$ satisfies
\begin{eqnarray}
\epsilon(K)=2(K^2-Q^2)+\frac{1}{\pi}\int_{-Q}^{Q}\left[
\frac{6|c|}{9c^2+4(K-K^{\prime})^2}\right. \nonumber \\
\left.+ \frac{|c|}{c^2+(K-K^{\prime})^2}-
\frac{2|c|}{4c^2+(K-K^{\prime})^2}\right]\epsilon(K^{\prime})dK^{\prime}.
\end{eqnarray}
Other gapless excitations such as the particle-hole and current
excitations can be expressed as the superposition of a single
particle and single hole excitations. The Fermi velocity is
\begin{equation}
v_F=\frac{\epsilon'(Q)}{\pi \rho_0(Q)}.
\end{equation}
At low temperatures $T\ll \Delta$, the spin degrees of freedom are
frozen completely. Thus the low temperature physics is almost the
same to that of the Lieb-Liniger model. As a Luttinger liquid, its
low-temperature specific heat and susceptibility behave as
\begin{equation}
C(T)=\frac{\pi^2}{3 v_F}T+o(T^2),{~~} \chi(T)\sim
e^{-\frac{c^2}{2T}},
\end{equation}
where we have taken the Boltzmann constant $k_B$ as our unit.

The general excited states are characterized by a set of real
$\{k_j\}$; a set of $k-\Lambda$ pairs described by Eqs.
(\ref{string1}-\ref{string2}) and $\Lambda$ $n$-strings taking the
form of $\Lambda_{\alpha,j}^{(n)}=\Lambda_{\alpha}^{(n)}+ i
(n+1-2j)c/2 +o(e^{-\delta L})$, where $j=1, \cdots, n$ and
$\alpha=1, \cdots, M_n$ with $n=1,2,\cdots$ and $M=\sum_n nM_n$.
Denote $\sigma^{\prime}$, $\rho$ and $\sigma_n$ as the densities
of bound pairs, real rapidities and $\Lambda$ $n$-strings in the
thermodynamic limit, respectively, and ${\sigma^{\prime}}^{h}$,
$\rho^{h}$ and $\sigma_n^{h}$  as the corresponding hole
densities. By minimizing the Gibbs free energy\cite{takahashi}, we
obtain the following coupled nonlinear integral equations
\begin{eqnarray}
&&\ln \eta^{\prime}
=2T^{-1}(k^2-\frac{c^2}{4}-\mu)-(a_5-a_1)*\ln(1+\xi^{-1}) \nonumber \\
&&\quad\quad -(a_6+a_4-a_2)*\ln(1+
{\eta^{\prime}}^{-1}),  \nonumber \\
&&\ln \xi =T^{-1}(k^2-\mu-2h)-(a_5-a_1)*\ln(1+
{\eta^{\prime}}^{-1}) \nonumber \\
&&\quad\quad +\ln \eta_1-(a_4+a_2+a_0)*\ln(1+\xi^{-1}), \nonumber \\
&&\ln \eta_1= G*[\ln(1+\eta_2)+\ln(1+\xi^{-1})],  \\
&&\ln \eta_n= G*[\ln(1+\eta_{n+1})+\ln(1+\eta_{n-1})], n=2,3,\cdots \nonumber \\
&&\lim_{n\rightarrow \infty} \frac{\ln \eta_n}{n}=\frac{h}{T},
\nonumber
\end{eqnarray}
where $a_n(x)=4n|c|/[\pi((nc)^2+(4x)^2)]$,
$\eta^{\prime}={\sigma^{\prime}}^{h}/\sigma^{\prime}$,
$\xi=\rho^{h}/\rho$, $\eta_n=\sigma_n^{h}/\sigma_n$,
$G(x)=c^{-1}sech(2\pi x/c)$, $f*g=\int f(x-y)g(y)dy$, and $h$ is the
external magnetic field. For $T=0$ and $h=0$, it is easily to deduce
that $\rho=\sigma_n=0$ and $\sigma^{\prime}\equiv\rho_0$. This also
confirms that the previously given ground state is the correct
ground state.
\par
For $c<0$, the interaction in $S=2$ channel is attractive while it
is repulsive in the $S=0$ channel. From the Bethe ansatz equations
we learn that the ground state is a incompressible ferromagnetic
state described by an $N$ string
\begin{eqnarray}
k_j= ic(N+1-2j),{~~~}j=1,2\cdots,N.
\end{eqnarray}

\section{Corresponding 3D model}

Now let us turn to the 3D case. An obvious fact is that two kinds
of condensation may occur in the corresponding 3D systems with
attractive interaction in the $S=0$ channel. One is the
conventional BEC and the other is the BCS like pair condensation
as indicated by the 1D exact result. An interesting question
arises: Is there any BCS-BEC crossover or BCS-BEC coexistence? To
answer this question, we consider the following Hamiltonian
\begin{equation}
H= -\sum_s \int a_s^{\dag}({\bf r}) \nabla^2 a_s({\bf r})d {\bf r} -
v \int p^{\dag}({\bf r}) p({\bf r}) d {\bf r}, \label{h3}
\end{equation}
where $p^{\dag}({\bf r})=[a_{\uparrow}^{\dag}({\bf r})
a_{\downarrow}^{\dag}({\bf r})+ a_{\downarrow}^{\dag}({\bf
r})a_{\uparrow}^{\dag}({\bf r})-a_{0}^{\dag}({\bf
r})a_{0}^{\dag}({\bf r})]/\sqrt{3}$ and $v$ is a positive coupling
constant. For simplicity, repulsive interaction in the $S=2$
channel is omitted since it is irrelevant to the pair
condensation. Motivated from the 1D exact result, we introduce the
order parameter of pair condensation as
\begin{equation}
{\cal O} =\left \langle V^{-1} \int p({\bf r}) d{\bf r}
\right\rangle_T,
\end{equation}
where $V$ and $\left\langle\cdots\right\rangle_T$ denote the
volume and the thermodynamic average, respectively. By using
BCS-like mean-field approximation, we linearize (\ref{h3}) as
\begin{eqnarray}
H&\approx&\sum_{{\bf k}}\left\{ \sum_s \epsilon({\bf k})
a_s^{\dag}({\bf k})a_s({\bf k}) -\frac{v{\cal O}}{\sqrt{3}}\left[
\sum_{\sigma} a_{\sigma}^{\dag}({\bf
k}) a_{\bar \sigma}^{\dag}({\bf - k}) \right.\right. \nonumber \\
&& \left.\left. -a_0^{\dag}({\bf k}) a_0^{\dag}(-{\bf k})+
h.c.\right] \right \} +Vv{\cal O}^2, \label{h6}
\end{eqnarray}
where $\epsilon({\bf k})={\bf k}^2$, $\sigma=\uparrow, \downarrow$
and $\bar \sigma $ means the spin flipped state. With the following
Bogoliubov transformations
\begin{eqnarray}
&&b_{\sigma}^{\dag}({\bf k}) = u({\bf k})a_{\sigma}^{\dag}({\bf k})
+ v({\bf k}) a_{\bar \sigma}(-{\bf k}), \nonumber \\
&&b_0^{\dag}({\bf k}) = u^{\prime}({\bf k})a_0^{\dag}({\bf k}) +
v^{\prime}({\bf k})a_{0}(-{\bf k}),
\end{eqnarray}
\begin{figure}[ht]
\begin{center}
\includegraphics[height=6cm,width=8cm]{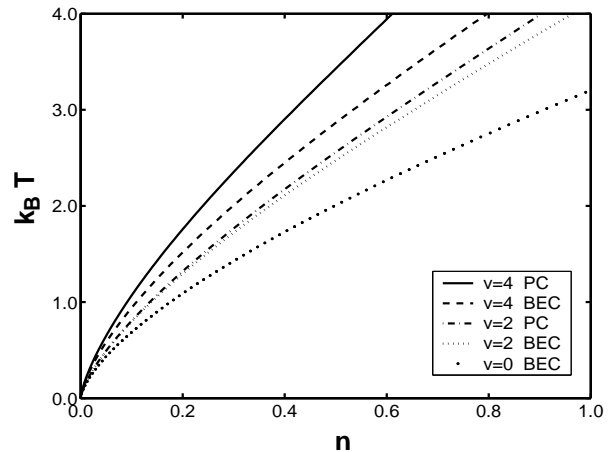}
\end{center}
\caption{The transiton temperatures $T_c^p$ and $T_c^b$ versus
density of particles $n$ for different interacting strength $v$.
The solid and dash-dot lines indicate $T_c^p$, while the dashed
and dotted lines indicate $T_c^b$. The asterisk line is $T_c^b$
for the ideal gas ($v=0$). $T_c^p>T_c^b$ for any positive $v$.}
\label{fig}
\end{figure}
where $u^2={u^{\prime}}^2=(g+1)/2$, $v^2={v^{\prime}}^2=(g-1)/2$,
$u^{\prime}v^{\prime}=-uv=g/f$, $g=f/\sqrt{f^2-4}$ and $f=
\sqrt{3}{\bf k}^2/(v{\cal O})$, the Hamiltonian (\ref{h6}) can be
diagonalized. The order parameter ${\cal O}$ and the chemical
potential $\mu$ are determined by the following self-consistent
equations
\begin{eqnarray}
\frac{1}{v} &=& \frac{1}{4\pi^2} \int_{0}^{\epsilon_F}
\frac{\sqrt{\epsilon}}{E}
\coth \frac{\beta E }{2} d \epsilon, \nonumber \\
n &=& \frac{1}{4\pi^2} \int_0^{\epsilon_F}
\frac{3\sqrt{\epsilon}}{2}\left(\frac{\epsilon-\mu}{E}\coth
\frac{\beta E}{2}-1\right) d\epsilon, \label{tc1}
\end{eqnarray}
where $\epsilon_F$ is the energy cutoff or band width in an optical
lattice, $E=\sqrt{(\epsilon - \mu)^2-4v^2{\cal O}^2/3}$, $n = \left
\langle V^{-1} \sum_s \int a_s^{\dag}({\bf r}) a_s({\bf r})d {\bf r}
\right \rangle_T$ is the density of particles, $\beta=1/T$. The
critical temperature $T_c^p$ for pair condensation is determined by
Eq. (\ref{tc1}) with ${\cal O}|_{T\to T_c^p}=0$. Interestingly,
besides the condensation of atom pairs, ordinary BEC also occurs at
low temperature $T<T_c^b$ when $\mu= 2v{\cal O}/\sqrt{3}$. The
numerical solutions of $T_c^b$ and $T_c^p$ for $v=2,4$ are depicted
in Fig. \ref{fig}. It is shown that both $T_c^b$ and $T_c^p$
increase with increasing interacting strength $v$ and the density of
particles $n$. Meanwhile, for a fixed $v$, $T_c^p$ is always larger
than $T_c^b$. When $v\to 0$, $T_c^p\to T_c^b$. Below $T_c^b$,
coexistence of pair condensation and BEC occurs. However, there is
no BCS-BEC crossover which usually occurs in fermion gases.

\section{Conclusion}

In conclusion, we propose an exactly solvable model describing the
low density limit of the spin-1 bosons in a 1D optical lattice.
Based on the exact result, a mean-field approach for the
corresponding 3D model is introduced. A new matter phase, i.e., the
pair condensate and coexistence with ordinary BEC are predicted. We
expect this new matter state could be realized in experiments.

\acknowledgments This work is supported by NSFC under grant
No.10474125, No.10574150 and the 973-project under grant
No.2006CB921300.

*Email: yupeng@aphy.iphy.ac.cn

\end{document}